\documentclass[aps,twocolumn,showpacs]{revtex4}
\usepackage{dcolumn}
\usepackage{float}
\usepackage{graphicx}
\usepackage{amsmath}
\usepackage{verbatim}
\usepackage{amssymb}
\usepackage{psfrag}
\usepackage{color}
\usepackage{bm}
\usepackage{wrapfig}
\usepackage{subfigure}
\usepackage{makeidx}
\usepackage{bm}
\usepackage{epsf}
\usepackage{multirow}
\usepackage{mathrsfs}
\usepackage{braket}
\usepackage[colorlinks,linkcolor=blue,urlcolor=blue,citecolor=blue]{hyperref}

\begin{document}
\title{Controllable Thouless Pumping Switching Dynamics of Gap Solitons Mediated by Finite
Bogoliubov Excitations}

\author{Tao Jiang$^{1}$}
\author{Jie Liu$^{2}$}

\author{Li-Chen Zhao$^{1,3,4,5}$}\email{zhaolichen3@nwu.edu.cn}

\address{$^{1}$School of Physics, Northwest University, Xi'an, 710127, China}
\address{$^{2}$Graduate School, China Academy of Engineering Physics, Beijing 100193, China}
\address{$^{3}$NSFC-SPTP Peng Huanwu Center for Fundamental Theory, Xi'an 710127, China}
\address{$^{4}$Shaanxi Key Laboratory for Theoretical Physics Frontiers, Xi'an 710127, China}
\address{$^{5}$Fundamental Discipline Research Center for Quantum Science and technology of Shaanxi Province, Xi'an 710127, China}
\begin{abstract}
We investigate the Thouless pumping dynamics of nonlinear gap solitons and attempt to realize
topological Chern number switching by modulating nonlinear parameters and varying the ramping
rate of the relative phase between periodic potentials. We find that gap solitons can undergo nonlinear instabilities accompanied by finite Bogoliubov excitations under near-adiabatic ramping. Such finite Bogoliubov excitations induce the particle loss of the solitons, leading to reversed propagation
directions that signals the occurrence of Chern number switching with analyzing the correspondence between soliton chemical potential and Bloch topological energy band. Our findings offer a feasible strategy for manipulating the Thouless pump dynamics of gap solitons mediated by finite Bogoliubov excitations, with implications for topological quantum transport and quantum computing applications.

\end{abstract}

\date{\today}
\maketitle

\section{Introduction}
Thouless pumping \cite{Thouless 1982, Thouless 1983, Thouless 1984} driven by periodic modulation provides a powerful probe of the underlying topological properties of energy bands and has been extensively investigated in a wide range of linear and even nonlinear systems \cite{Kraus 2012,Verbin 2015,Ke 2016,Cerjan 2020,Wang 2022,Lohse 2016,Nakajima 2016,Lu 2016,Nakajima 2021,Walter 2023, Tangpanitanon 2016,Jurgensen 2021,Jurgensen 2022,Ye Fangwei 2022,Ye Fangwei2 2022,Jurgensen 2023,Citro 2023}.  In linear systems, quantized pumping reflects the displacement of the Wannier-state center of mass (COM), and the transport distance per cycle remains invariant throughout the adiabatic evolutions \cite{Resta 2007}.
In nonlinear systems, the transport of stable gap solitons \cite{Sterke 1994, Chen 1987, Louis 2003,Kartashov 2011} also remain unchanged \cite{Gong Jiangbin 2023,Z. Liang 2024,Xu Yong1 2025,Xu Yong2 2025,Mostaan 2022,Julius 2026} under the usual adiabatic driving. In adiabatic evolution,  a system prepared in an initial eigenstate continuously follows the corresponding instantaneous eigenstate throughout the pumping cycle,  is a key prerequisite for realizing a quantum pumping described by topology of energy bands \cite{Thouless 1983, Thouless 1984}. It is usual to term the operations for adiabatic evolution holding in linear systems as adiabatic operations. The pumping displacement per cycle remains unchanged under the adiabatic operations even in nonlinear systems in previous studies  \cite{Ye Fangwei 2022, Jurgensen 2021}, and the pumping quantities are described by related topological invariants. These results form a conventional perspective on adiabatic quantum transport.

In this work, we demonstrate that gap solitons exhibit some spontaneous transitions of quantum pumping under the usual adiabatic driving operations, which is termed as a ``pump switching" phenomenon beyond the above conventional perspective on adiabatic quantum transport. Our analysis indicates that the solitons exhibit weak instability windows within a driving cycle. The limited instability window provides possibilities to realize finite Bogoliubov excitations in the driving process, in contrast to the soliton breaking in strongly unstable cases mentioned in previous studies \cite{Julius 2026,Jurgensen 2022}. The finite Bogoliubov excitations induce the quantum pump transition of solitons, accompanying the soliton energy level (SEL) transition. We further show that the transport switching can be controlled by adjusting the driving speed or varying the initial soliton particle number. Soliton cascade transitions and stimulated transitions by colliding with solitons are also demonstrated. These findings uncover the correlation between Bogoliubov excitations, SEL transitions and quantum pumping dynamics, enriching the theoretical framework for nonlinear quantum wave systems and Bloch-band-derived solitons under periodic driving.


\section{Thouless pumping switching of Solitons}
\label{Sec2}
\begin{figure}[htbp]
	\includegraphics[width=0.5\textwidth]{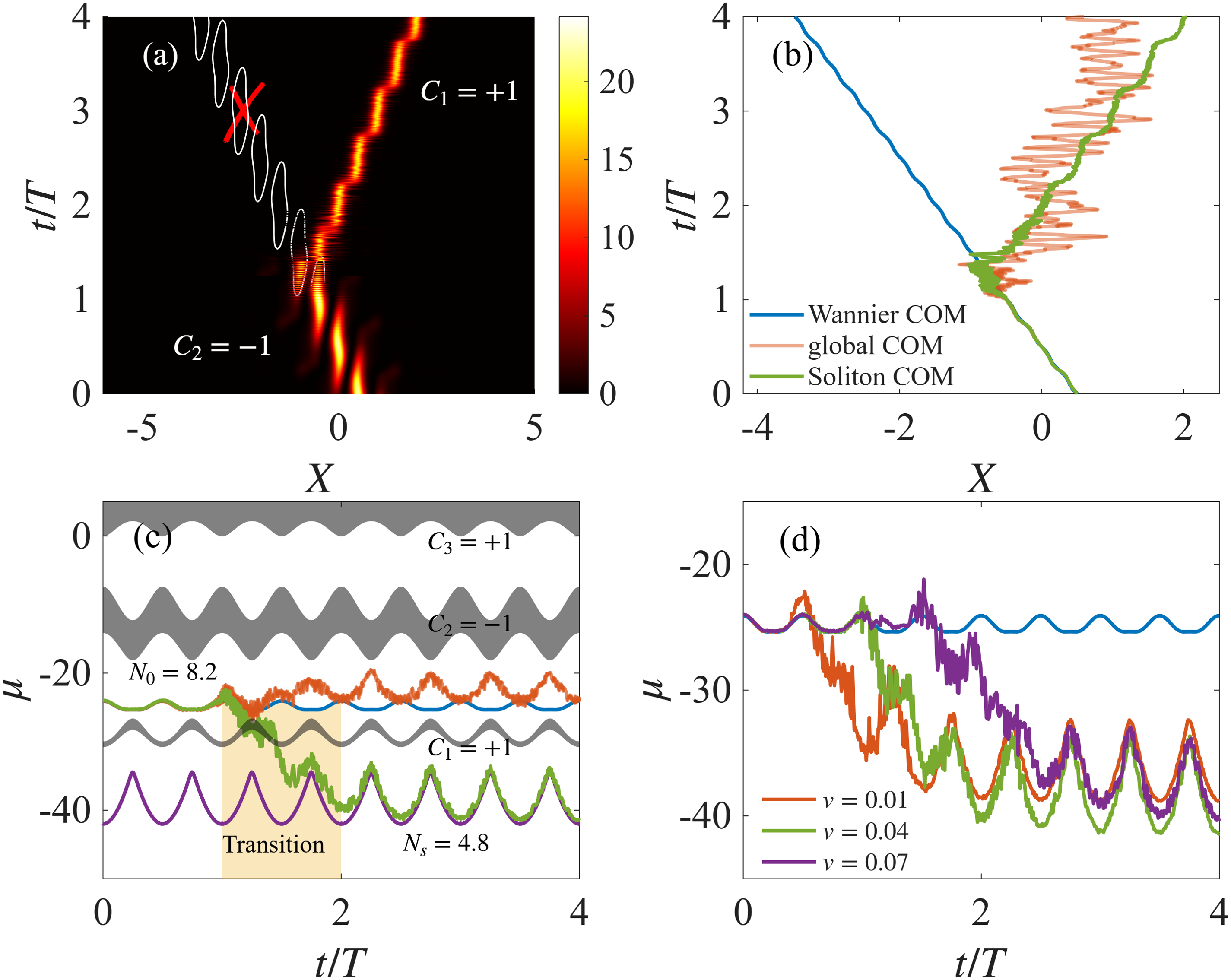}
	\caption{(a) Transport switching of a gap soliton ($N_0=8.2$, $v=0.05$). White contours indicate the instantaneous soliton eigenstates.
		(b) COM trajectories of the soliton (green), linear Wannier state (blue), and full wavefunction (red).(c) SEL transition. Gray shaded regions denote the Bloch bands; blue and purple curves show the SELs for $N_0=8.2$ and $N_s=4.8$, respectively. Red and green curves represent $\bar{\mu}(t)$ and $\bar{\mu}_s(t)$, obtained from the global wavefunction and the local soliton, respectively. (d) Evolution of $\bar{\mu}_s(t)$ for different driving speeds.}
	\label{Fig.1}
\end{figure}
We consider a Bose-Einstein condensate loaded into a one-dimensional optical superlattice, to investigate the motion of solitons under the periodic driving operations. Within the mean-field approximation, its dynamics can be described by the dimensionless Gross-Pitaevskii (GP) equation\cite{Gross 1961, Pitaevskii 1961}.
\begin{eqnarray}
i\partial _t\psi =-\frac{1}{2}\partial _{x}^{2}\psi +g\left| \psi \right|^2\psi +V\left( x,t \right) \psi,
\end{eqnarray}
where $\psi$ is the condensate wave function, and $g$ characterizes the interatomic interaction strength. The potential $V(x,t)=D_1\cos ^2\left( \pi x/a_1 \right) +D_2\cos ^2\left( \pi x/a_2+vt \right) $ represents an optical superlattice, where the first term corresponds to a stationary primary lattice, while the second term describes a moving lattice. The $D_1$ and $D_2$ denote the depths of the two lattice potentials, $a_1$ and $a_2$ are their spatial periods. The parameter $v$ denotes the phase sweep rate, and the corresponding driving period is $T=\pi/v$ Unless otherwise specified, we set $D_1=D_2=25$, $a_1=1/2$, and $a_2=1$ throughout this work. This optical superlattice potential provides a standard platform for studying soliton Thouless pumping.  For attractive interactions ($g<0$) and sufficiently small driving speeds, gap solitons bifurcating from the lower edge of the first Bloch band exhibit both integer and fractional quantized transport~\cite{Ye Fangwei 2022}. Similar discussions  can be made for solitons with repulsive interactions \cite{YP Zhang 2025}. In general, the quantized transport of solitons can still be predicted from the Chern number $C_n$ of the Bloch band (with $n$ the band index) that the soliton predominantly occupies~\cite{Ye Fangwei 2022,Jurgensen 2022,Julius 2026}. It is worth noting that these solitons do not exhibit spontaneous transport switching under the usual sweeping rates satisfying the adiabatic condition for the systems with $g=0$. However, we uncover a striking pump switching phenomenon, namely a reversal of the transport direction, by investigating the transport of gap solitons bifurcating from higher Bloch bands under varying driving rates within the adiabatic regime for the corresponding linear systems.

We take gap solitons bifurcating from the second Bloch band ($C_2=-1$) as an example to illustrate the pump switching. Our numerical simulations show that the soliton is pumped by one unit cell along the negative direction of the lattice during the first driving cycle, consistent with conventional soliton transport~\cite{Jurgensen 2021}. Interestingly, the soliton suddenly reverses its propagation direction after one cycle (see Fig.~\ref{Fig.1}(a)), in contrast to  continued leftward transport along the white contour expected by the previous studies. After the transport switching, the soliton predominantly occupies the first Bloch band ($C_1=+1$). Such spontaneous pump switching does not occur in linear systems, since the driving speeds lie within the adiabatic regime of the linear system. Its Wannier COM dynamics simply correspond to a repetition of the first cycle motion, as shown by the blue curve in Fig.~\ref{Fig.1}(b). By contrast, the COM of the full wavefunction (see the red curve in Fig.~\ref{Fig.1}(b)) develops strong fluctuations after the switching due to dispersive wave radiation. To eliminate this effect, we introduce absorbing boundary conditions to filter out most dispersive waves affecting soliton's COM by periodic boundaries. The soliton peak transport was used to characterize the pump for single peak soltions when there were dispersive wave radiation \cite{Gong Jiangbin 2023}.  The peak identification scheme does not work well for solitons with multiple peaks. Therefore, we define the soliton COM by integrating the particle density associated with the soliton wave function $\psi_s(x,t)$ within a finite spatial window. The local spatial window, denoted by $\Omega_s$, is chosen to be five times the full width at half maximum of the soliton. The resulting trajectory accurately reproduces the transport dynamics with greatly reduced fluctuations (see the green curve in Fig.~\ref{Fig.1}(b)), indicating that quantities related to transport switching should be defined using the local soliton wavefunction $\psi_s(x,t)$ rather than the global wavefunction $\psi(x,t)$.

The switching of soliton transport indicate that soliton transport involves some non-adiabatic dynamics. Since non-adiabatic evolution in linear quantum systems is typically accompanied by coupling and transitions between instantaneous eigenlevels, it is natural to seek an analogous description for soliton transport. The system eigenvalues $\mu$ are usually obtained by self-consistently solving the nonlinear eigenproblem $\hat{H}(\psi)\psi=\mu\psi$ at a fixed total particle number $N_0 =\int |\psi(x,t)|^2 dx$, which contains both soliton's particle number and dispersive waves's. The SELs $\mu_s$ are defined from the same nonlinear eigenvalue problem, but only taking the localized  particle number of the soliton $N_s=\int_{\Omega_s} |\psi_s(x,t)|^2 dx$ ($\Omega_s$ is also chosen to be five times the full width at half maximum of the soliton).  We compute the system eigenlevels for the scenario shown in Fig.~\ref{Fig.1}(a), including the linear Bloch bands and the SEL corresponding to an initial particle number $N_0=8.2$, represented by the gray region and the blue curve in Fig.~\ref{Fig.1}(c), respectively. Due to the weak nonlinearity having only a minor effect on the bulk bands, the bandgap structure is preserved \cite{Louis 2003, Kartashov 2011}, and we mark the nonlinear Bloch bands by linear Bloch bands.  For comparison with the SELs, we calculate the instantaneous expectation value of the GP Hamiltonian with respect to the global wavefunction $\psi(x,t)$, namely, $\bar{\mu}(t)=\frac{\int \psi^{*}(x,t)\hat{H}[\psi(x,t)]\psi(x,t)dx}{\int |\psi(x,t)|^2 dx}$, as well as the corresponding instantaneous expectation value $\bar{\mu}_s(t)$ evaluated using the local soliton wavefunction $\psi_s(x,t)$. These quantities are shown as the red and green curves in Fig.~\ref{Fig.1}(c), respectively. During the first transport cycle, $\bar{\mu}(t)$ and $\bar{\mu}_s(t)$  both coincide with the initial SEL and system eigenvalue, indicating adiabatic evolution. After passing through the yellow region, they deviate from each other.  The $\bar{\mu}(t)$ remains in the original bandgap without matching any SEL and even the system eigenvalue, which is inconsistent with the stable soliton transport after switching. But $\bar{\mu}_s(t)$ undergoes a short oscillatory stage and then gradually stabilizes within the semi-infinite bandgap, where it coincides with another SEL determined by the stable soliton's particle number $N_s = 4.8$ (see the purple curve in Fig.~\ref{Fig.1}(c)). Thus the introduction and definition of SEL are necessary and reasonable. We refer to this dynamical process as a spontaneous SEL transition.

Moreover, the transition result depends on the sweep rate, with slower driving leading to an earlier occurrence of the transition, while the post-transition SEL generally differs in its position, as shown in Fig. \ref{Fig.1}(d). Unlike conventional quantum mechanical level transitions, this process occurs over a finite timescale.  The final stable soliton's particle number varies with changing the sweeping rate. These characters are only possessed by nonlinear systems, and they could be induced by the interplay among nonlinear instability, soliton profiles, external potentials,  and the sweeping rates. In the following section, we further elucidate the physical mechanism underlying this spontaneous process.

\section{The underlying mechanism for pumping switching}
\label{Sec3}

\begin{figure}[htbp]
	\includegraphics[width=0.5\textwidth]{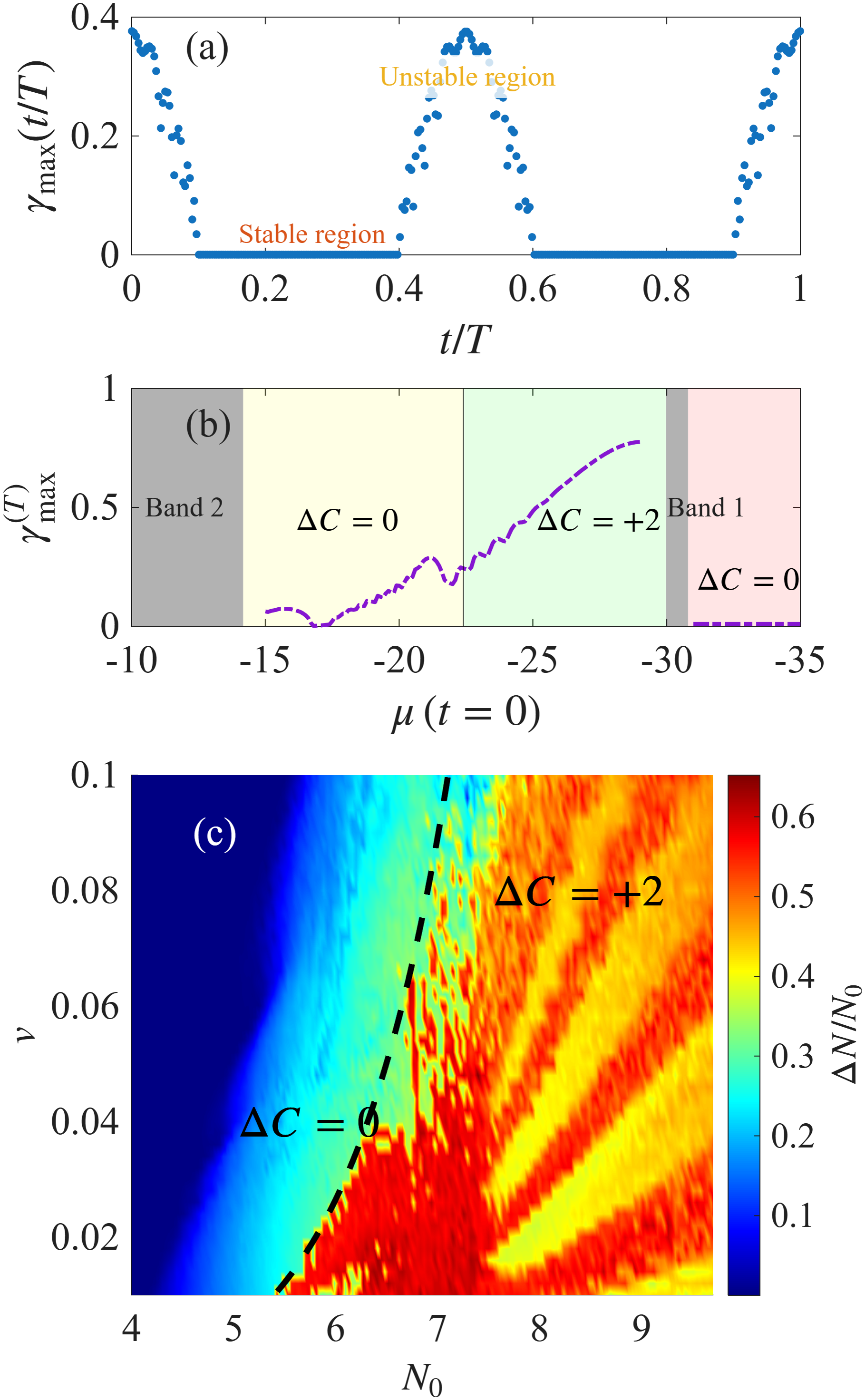}
	\caption{(a) Instantaneous stability phase diagram of the soliton over one driving cycle for $N_0=8.2$.
		(b) Stability phase diagram of the SELs within the bandgap. For $v=0.05$, no transport switching occurs in the yellow and red regions over five driving cycles, whereas transport switching occurs in the green region. The purple curve indicates the maximum growth factor over one driving cycle.
		(c) Phase diagram of transport switching over five driving cycles as functions of the initial particle number and driving speed. The black dashed curve indicates the approximate phase boundary.}
	\label{Fig.2}
\end{figure}
The above mismatch of $\bar{\mu}(t)$ and SELs and even the system eigenvalue, suggests that the dispersive wave radiation  involves  Bogoliubov excitations going beyond the description of condensate dynamics. Namely, the particle number in the condensate decrease by radiating dispersive waves. To elucidate its microscopic origin, we perform a Bogoliubov-de Gennes (BdG) analysis \cite{Bogoliubov 1946, Pethick 2008,J. Yang 2010} on the instantaneous soliton states within a single driving period. By solving the BdG eigenvalue problem, we obtain a set of quasiparticle excitation frequencies. When the eigenfrequency $\omega$ acquires a nonzero imaginary part, i.e., $\mathrm{Im}[\omega] = \gamma \neq 0$, any perturbation is exponentially amplified, indicating dynamical instability of the soliton state in the corresponding parameter regime. Here, $\gamma$ characterizes the growth rate of the unstable modes. We focus on the maximal instantaneous BdG growth rate $\gamma_{\max}(t/T)$ and plot its evolution over one driving period in Fig.~\ref{Fig.2}(a). We find that the instantaneous soliton state is not stable throughout the entire driving cycle, but instead alternates between stable and weakly unstable regimes. Notably, the stable regions suppress the rapid growth of unstable modes, allowing coherent soliton transport over long timescales, whereas the unstable regions provide some limited windows for Bogoliubov excitations. These weak instability windows provide possibilities to realize finite Bogoliubov excitations by changing the sweep rates.

We further perform the BdG analysis of instantaneous eigenstates with different particle numbers in both the first bandgap and the semi-infinite bandgap to see why the soliton bifurcating from the first band does not exhibit any pump switching. We evaluate the maximal growth rate over a single driving period, i.e. $\gamma_{\max}^{(T)} = \max_{t/T \in [0,1]} \gamma_{\max}(t/T)$, and plot it as a function of the soliton eigenvalue at $t=0$, together with the Bloch bands, as shown in Fig.~\ref{Fig.2}(b). The results show that solitons in the first bandgap exhibit instability windows within one driving period, with $\gamma_{\max}^{(T)}$ increasing toward the first Bloch band, indicating reduced stability. We emphasize that the alternation of stable and unstable windows is crucial for soliton eigenlevel transitions, in contrast to strongly unstable dynamics where persistent instability leads to rapid soliton breakup and prevents adiabatic following of any eigenlevel \cite{Julius 2026, Jurgensen 2022}. Weak and intermittent instability instead produces finite Bogoliubov excitations, allowing relaxation toward nearby stable states and thereby enabling eigenlevel transitions. Transport switching depends on the driving speed of the system, which determines the SEL on which the soliton settles after the transition. If the soliton remains in the first bandgap after undergoing finite Bogoliubov excitations, transport switching does not occur. To quantify this, we simulate transport over five driving periods at $v = 0.05$ and define switching via the Bloch band Chern number difference $\Delta C$, where $\Delta C = 0$ indicates no switching and $\Delta C \neq 0$ indicates switching. We find that the yellow region in Fig.~\ref{Fig.2}(b) shows no switching while the green region does, although both exhibit SEL transitions. In contrast, solitons in the red region (semi-infinite bandgap) remain dynamically stable and do not undergo spontaneous eigenlevel transitions; most previous studies focus on this regime \cite{Ye Fangwei 2022}, where no switching occurs.

The sweep rate determines the residence time of the soliton in the instability window. Meanwhile, SELs associated with different particle numbers possess distinct instability windows. For example, the initial soliton in Fig. \ref{Fig.1} exhibits some limited instability windows, but the final stable solitons in Fig. \ref{Fig.1} all exhibit no instability windows during each sweeping period. Consequently, transport switching can be controlled through both the initial particle number and the driving speed. A phase diagram of soliton transport over five driving periods as a function of the driving speed and initial particle number is presented in Fig.~\ref{Fig.2}(c), where $\Delta N/N_0$ denotes the ratio of the particle loss after the transition to the initial particle number, showing that larger particle numbers and slower driving both favor the occurrence of transport switching. The black curve approximately separates the regimes with and without transport switching.

\section{Soliton Cascade Transitions and Stimulated Transitions}
\label{Sec4}
\subsection{Soliton Cascade Transitions}
\begin{figure}[htbp]
	\includegraphics[width=0.5\textwidth]{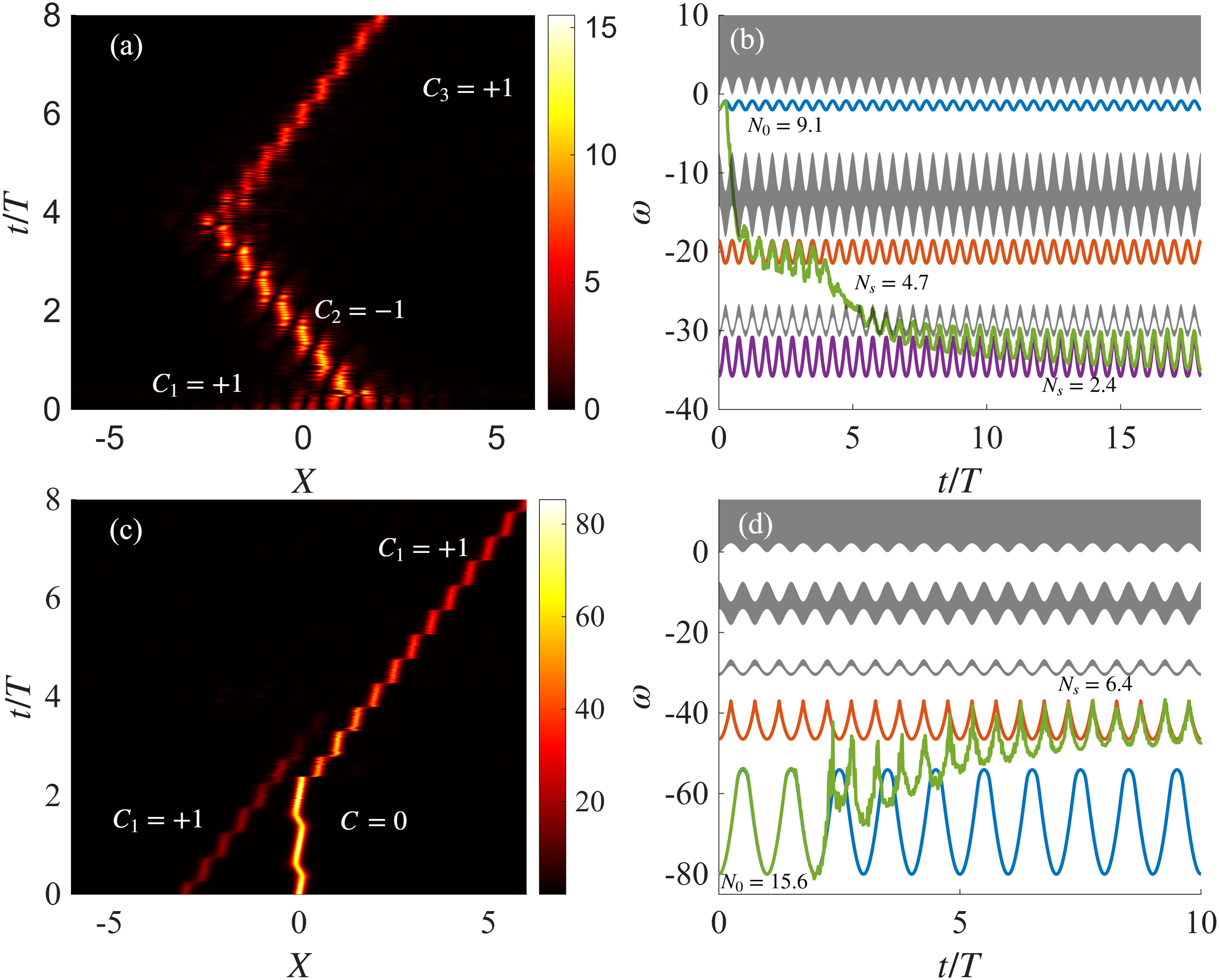}
	\caption{(a) Dynamical evolution of a soliton undergoing cascading SEL transitions. (b) Cascading SEL transitions. The blue curve corresponds to the initial SEL ($N_0=9.1$), while the red and yellow curves correspond to the SELs after the first and second transitions, with particle numbers $N_0=4.7$ and $N_0=2.4$, respectively.(c) Schematic illustration of a stimulated SEL transition, showing a transporting soliton (left) and a trapped soliton (right). (d) Collision induced SEL transition. The blue and red curves correspond to the SELs before and after the transition ($N_0=15.6$), respectively. The green curve shows the instantaneous expectation value $\bar{\mu}_s(t)$.}
	\label{Fig.4}
\end{figure}

The transport switching of the soliton fundamentally originates from SEL transitions induced by finite Bogoliubov excitations. Similar to conventional quantum transitions, SEL transitions may involve more than two eigenlevels. Because different SELs exhibit distinct stability properties, nonlinear instabilities can persist after a transition, leading to successive transitions among multiple SELs and resulting in a spontaneous cascading transition process. We demonstrate the cascading transition using a soliton ($N_0$ = 9.1) prepared in the third Bloch band (C = +1). Under this parameter condition, the unstable region spans a larger portion of a single driving period, while the instability is stronger, resulting in an earlier onset of transport switching. After the switching event, the soliton propagates approximately four cycles along the negative spatial direction before its transport direction reverses again, as shown in Fig.~\ref{Fig.4}(a). By analyzing the expectation value $\bar{\mu}_s(t)$, one can clearly observe that the soliton first undergoes a rapid transition from the initial SEL to a red SEL in the first bandgap, followed by adiabatic following and stable transport along this eigenlevel for a finite duration. It subsequently undergoes another transition into the semi-infinite bandgap, and after a relaxation process, finally stabilizes on the purple SEL.

\subsection{Stimulated Transitions}
The preceding discussion focused on spontaneous SEL transitions arising from finite Bogoliubov excitations of gap solitons. Here, we demonstrate that such transitions can also be externally induced, leading to a change in the transport dynamics. We prepare a soliton with particle number $N_0=15.6$ in the semi-infinite bandgap. Its corresponding SEL (blue curve in Fig.~\ref{Fig.4}(d)) remains linearly stable throughout the driving cycle. Owing to its large particle number, the soliton remains trapped and exhibits no transport, corresponding to a multiband Chern number of $C=0$~\cite{Ye Fangwei 2022}.
We then prepare a second soliton with particle number $N_0=4.8$, occupying only the lowest Bloch band ($C=+1$), far from the trapped soliton, as shown by the left soliton in Fig.~\ref{Fig.4}(c). Since the low-particle-number soliton undergoes normal transport, it collides with the trapped soliton after approximately two driving periods. The collision-induced particle loss enables the originally trapped soliton to overcome self-trapping and evolve along a higher SEL with $N_s=6.4$ (red curve in Fig.~\ref{Fig.4}(d)). Meanwhile, owing to its small initial particle number, the left soliton is destroyed during the collision, with most of its particles dispersing into linear waves.

\section{conclusion and discussion}
\label{Sec5}

This work breaks the conventional understanding of adiabatic quantum pumping by revealing that finite Bogoliubov excitations can trigger quantum pump transitions of gap solitons in nonlinear systems, even when the corresponding linear systems operate adiabatically. It identifies a pump switching phenomenon originating from the finite Bogoliubov excitations induced by the limited instability windows in a driving cycle, which differs fundamentally from strong instability scenarios in prior research. SELs can describe and explain the pumping switching behaviors. The findings uncover the correlation between Bogoliubov excitations, SEL transitions and quantum pumping dynamics.

The controllable pump switching behavior modulated by driving speed and initial particle number of solitons provides feasible strategies for manipulating soliton dynamics. These results offer theoretical support for the design of quantum pumping devices, all-optical switching elements and nonlinear quantum signal processors. They also provide a new  perspective for the development of quantum transport control technologies in optical lattices, Bose-Einstein condensates and other nonlinear quantum platforms.

\section*{Acknowledgments}
This work is supported by the National Natural Science Foundation of China (Contract No. 12375005, 12235007, 12247103).

%
%
%
%
%
%
%


\end{document}